\documentclass{aastex}
\usepackage{epsfig}
\begin{document}
\newcommand{\beq}{\begin{equation}}
\newcommand{\eeq}{\end{equation}}
\newcommand{\ber}{\begin{eqnarray}}
\newcommand{\eer}{\end{eqnarray}}
\newcommand{\ld}{\Lambda}
\newcommand{\m}{{\rm m}}
\newcommand{\om}{\Omega_{0\rm m}}
\newcommand{\oml}{\Omega_l}
\newcommand{\omx}{\Omega_{\Lambda_{\rm b}}}
\newcommand{\lleq}{\lower0.9ex\hbox{ $\buildrel < \over \sim$} ~}
\newcommand{\ggeq}{\lower0.9ex\hbox{ $\buildrel > \over \sim$} ~}\newcommand\del{\delta}
\newcommand{\ddel}{\delta^{\prime}}
\newcommand{\dd}{\delta^{\prime}_0}

\newcommand{\etal}{{\it et al. }}
\newcommand{\ie}{i.e.~}
\newcommand{\etc}{e.t.c.~}
\newcommand{\eg}{e.g.~}
\newcommand{\n}{\noindent}

\newcommand{\aphj}{Astroph.~J.~}
\newcommand{\mn}{Mon.~Not.~Roy.~Ast.~Soc.~}
\newcommand{\asta}{Astron.~Astrophys.~}
\newcommand{\astj}{Astron.~J.~}
\newcommand{\pl}{Phys.~Rev.~Lett.~}
\newcommand{\pd}{Phys.~Rev.~D~}
\newcommand{\nucp}{Nucl.~Phys.~}
\newcommand{\natr}{Nature~}
\newcommand{\plb}{Phys.~Lett.~B~}
\newcommand{\jetpl}{JETP ~Lett.~}
\newcommand{\jcaph}{J.~Cosmol.~Astropart.~Phys.~}

\title{Reconstructing Cosmological Matter Perturbations using Standard 
Candles and Rulers}

\shorttitle{Reconstructing Cosmological Matter Perturbations} 
\shortauthors{U. Alam, V. Sahni, A. A. Starobinsky}

\author{Ujjaini Alam}
\affil{ISR-1, ISR Division, Los Alamos National Laboratory, Los 
Alamos, NM 87545, USA}
\email{ujjaini@lanl.gov}
\author{Varun Sahni}
\affil{Inter-University Centre for Astronomy and Astrophysics, 
Pune 411~007, India}
\email{varun@iucaa.ernet.in}
\author{Alexei A. Starobinsky}
\affil{Landau Institute for Theoretical Physics, Moscow 119334, 
Russia ;}
\affil{RESCEU, Graduate School of Science, The University
of Tokyo, Tokyo 113-0033, Japan}
\email{alstar@landau.ac.ru}

\thispagestyle{empty}

\sloppy

\begin{abstract}
\small{ 
For a large class of dark energy (DE) models, for which the effective
gravitational constant is a constant and there is no direct exchange
of energy between DE and dark matter (DM), knowledge of the expansion
history suffices to reconstruct the growth factor of linearized
density perturbations in the non-relativistic matter component on
scales much smaller than the Hubble distance. In this paper we develop
a non-parametric method for extracting information about the
perturbative growth factor from data pertaining to the luminosity or
angular size distances. A comparison of the reconstructed density
contrast with observations of large scale structure and gravitational
lensing can help distinguish DE models such as the cosmological
constant and quintessence from models based on modified gravity
theories as well as models in which DE and DM are either unified, or
interact directly. We show that for current SNe data, the linear
growth factor at $z=0.3$ can be constrained to $5 \%$, and the linear
growth rate to $6\%$.  With future SNe data, such as expected from the
JDEM mission, we may be able to constrain the growth factor to $2-3\%$
and the growth rate to $3-4 \%$ at $z=0.3$ with this unbiased,
model-independent reconstruction method. For future BAO data which
would deliver measurements of both the angular diameter distance and
Hubble parameter, it should be possible to constrain the growth factor
at $z=2.5$ to $9\%$. These constraints grow tighter with the errors on
the datasets.  With a large quantity of data expected in the next few
years, this method can emerge as a competitive tool for distinguishing
between different models of dark energy.  }
\end{abstract}


\maketitle

\section{Introduction}\label{intro}

Over the last decade, observations of Type Ia supernovae have shown
that the expansion of the universe is currently accelerating
\citep{accl,accl1,accl2,accl3,accl4,accl5,accl6,accl7,union}. This
remarkable discovery has led cosmologists to hypothesize the presence
of dark energy (DE), a negative pressure energy component which
dominates the energy content of the universe at present. Many theories
have been propounded to explain this phenomenon, the simplest of which
is the cosmological constant $\ld$, with a constant energy density and
the equation of state $w=-1$. Although $\ld$ appears to explain all
current observations satisfactorily, to do so its value must
necessarily be very small $\ld/8\pi G \simeq 10^{-47}$GeV$^4$. So, it
represents a new small constant of nature in addition to those known
from elementary particle physics, many of them being very small if
expressed in the Planck units.  However, since it is not known at
present how to derive $\ld$ from these small constants and it is also
unclear if DE is in fact time independent, other phenomenological
explanations for cosmic acceleration have been suggested \citep[see
  reviews][]{DE_review,rev2,rev3,rev4,rev5,rev6,ss06}. These are based
either on the introduction of new physical fields (quintessence
models, Chaplygin gas, etc.), or on modifying the laws of gravity and
therefore the geometry of the universe (scalar-tensor gravity, $f(R)$
gravity, higher dimensional `Braneworld' models \etc). The plethora of
competing dark energy models has led to the development of parametric
and non-parametric methods as a means of obtaining model independent
information about the nature of dark energy directly from observations
\citep [see][and references
  therein]{star98,DE_recon,recon1,recon2,recon3,recon4,recon5,recon6,recon7,recon8,recon9,sss08,ss06}.

The next decade will see the emergence of many new cosmological
probes. A large number of these are likely to make important
contributions to the field of dark energy. The Sloan Digital Sky
Survey began its stage III observations in 2008, and its Baryon
Oscillation Spectroscopic Survey (BOSS) is expected to map the spatial
distribution of luminous galaxies and quasars and detect the
characteristic scale imprinted by baryon acoustic oscillations in the
early universe \citep{sdss}. The Joint Dark Energy Mission (JDEM) is
expected to discover a large number of supernovae, and also provide
important data on weak-lensing and baryon acoustic oscillations
\citep{jdem}. The Square Kilometer Array (SKA) will map out over a
billion galaxies to redshift of about 1.5, and is expected to
determine the power spectrum of dark matter fluctuations as well as
its growth as a function of cosmic epoch \citep{ska}. Important clues
to the growth of structure will also come from current and future weak
lensing surveys (CFHTLS, DES, JDEM, EUCLID, SKA, LSST), galaxy
redshift-space distortions \citep{guzo,percival, percival2} as well as
galaxy cluster mass functions at different redshifts $z$ \citep{vikh,
  rap}. With the wealth of data expected to arrive over the next
several years, it is important to explore different methods of
analyzing these datasets in order to extract the optimum amount of
information from them. In this paper we explore the possibility of
reconstructing the linearized growth rate of density perturbations in
the non-relativistic matter component, $\delta(z)$, taken at some
fixed comoving scale much less than the Hubble distance, from datasets
which have traditionally been used to explore only the smooth
background universe, \eg luminosity distance and angular diameter
distance data.

In the case of {\em physical} DE, the effective gravitational constant
appearing in the equation for linear density perturbations in the
matter component coincides with the Newton gravitational constant $G$
measured in the laboratory and using Solar system tests. If,
additionally, there is no direct non-gravitational interaction between
DE and DM in the physical reference frame, so that the DE
energy-momentum tensor is covariantly conserved, the density contrast
reconstructed in this manner should match that determined directly
from observations of large scale structure. In this case the methods
developed in this paper will provide an important consistency check on
DE models such as $\ld$ and Quintessence. On the other hand, {\em
  geometrical} models of DE (Braneworlds, scalar-tensor gravity, etc.)
usually predict a different growth rate for $\delta(z)$ from that in
general relativity (GR). Models where DE has a direct
non-gravitational interaction with DM, or where DE and DM are unified,
have a similar property even in the framework of GR. In this case, a
reconstruction of the linearized density contrast from observations of
standard candles/rulers will not match with $\delta$ determined
directly from large scale structure
\citep{f_dgp,bert,ishak,knox,chiba,MMG,haiman,hu,pol,koy}. Currently
reconstructed values of the growth rate from galaxy redshift
distortions \citep{guzo,sdss2,sdss2-1,slaq} are not very constrictive,
but future missions like Euclid \citep{euclid} are expected to
constrain the growth rate tightly. Therefore comparing the results
from future supernova data, using the methods described in this paper,
to those from future large scale structure data will help address
important issues concerning the nature of gravity and dark energy.

This paper is organized as follows. In section~\ref{meth}, we describe
the reconstruction technique and the data used to test this
method. Section~\ref{res} shows the results and examines the
dependence of the method on various factors such as the redshift
distribution of the data and information on other cosmological
parameters. The conclusions are presented in section~\ref{concl}.

\section{Methodology}\label{meth}

In the longitudinal (quasi-Newtonian) gauge, the perturbed, spatially
flat, Friedman-Robertson-Walker (FRW) metric is defined by the line
element
\beq
ds^2 = -(1+2\phi)dt^2 + (1-2\psi)a^2(t)d{\vec x}^2~,
\eeq
where $\phi = \psi$ in GR if matter is free of anisotropic stresses 
(we assume DM to be cold and neglect small effects from the neutrino
component which produces $\phi\not=\psi$). The Newtonian potential 
$\phi$ and the non-relativistic matter density contrast 
\beq
\delta_m = \frac{\rho_m({\vec x}, t) - \bar{\rho}(t)}{\bar{\rho}(t)},
\eeq
are linked via the linearized Poisson equation
\beq\label{eq:Poisson}
k^2\phi = -4\pi Ga^2\rho_m\delta_m~.
\eeq

If the DE energy-momentum tensor is covariantly conserved, then
$\rho_m\propto a^{-3}$. In this case it is straightforward to show
\citep[see, \eg][]{growth,star98} that on scales much smaller than the
effective Jeans scale for DE, $\lambda_J\sim c_sH^{-1}$, where $H(z)
\equiv{\dot a}/a$ is the Hubble parameter and $c_s$ is the effective
DE sound velocity ($c_s=1$ for standard quintessence), linearized
matter density perturbations in a FRW universe containing DE with an
arbitrary effective equation of state $w(t)\equiv p_{DE}/\rho_{DE}$
satisfy the same equation as in the case of a standard FRW model
driven by dust and a cosmological constant 
\citep{peebles}:
\beq\label{eq:diff}
\ddot{\delta}_m+2H\dot{\delta}_m - 4\pi G \rho_m \delta_m = 0 \,\,,
\eeq
(we ignore the subscript in $\delta_m$ in the ensuing discussion). 
However, the generic textbook solution \citep{peebles,coles}
\beq\label{stand-sol}
\delta \propto H(z)\int_{z_0}^z \frac {1+z_1}{H^3(z_1)} \ dz_1
\eeq 
is not applicable now, apart from the following cases: dust-like
matter, a non-zero spatial curvature (and/or a tangled network of
cosmic strings) and a cosmological constant, for which
$H^2(z)=C_1+C_2(1+z)^2+C_3(1+z)^3$. The same refers to the other
well-known expression valid only for dust and a cosmological constant:
\beq\label{stand-sol1}
\delta \propto a(t) - H(t)\int^t a(t_1)\,dt_1~,
\eeq
 see e.g. \citep{KS85,bert}.\footnote{The
expression (\ref{stand-sol1}) is, in fact, the first term in the
long-wave (super-Hubble) expansion of the adiabatic mode of a
comoving density perturbation if the perturbed matter pressure
tensor is proportional to the unit one and the spatial curvature
may be neglected. Then $\phi=\psi=-\zeta(\vec x)\left(
1-\frac{H}{a}\int_{t_1}^t a \, dt\right)$, where the
gauge-invariant curvature perturbation $\zeta$ does not depend on
time for the growing adiabatic mode, see e.g. \citep{PS92,bert}.
The quantity $t_1$ is free and may be chosen to coincide with the
moment of the first Hubble radius crossing during inflation
(another choice would correspond to adding a decaying adiabatic
mode with an arbitrary amplitude). Then, using Eq. (\ref{eq:Poisson})
and the fact that $\rho_m\propto a^{-3}$, the formula
(\ref{stand-sol1}) follows. Since DE is practically unclustered at
sub-Hubble scales, it is tempting to try to use this formula for
$\lambda_J\ll \lambda \ll H^{-1}$, too. However, as pointed above,
this works only if DE is a cosmological constant.} 
Thus, for an arbitrary physical DE, Eq~(\ref{eq:diff}) has to be solved
numerically. Since there are no terms depending on the
perturbation wave vector ${\bf k}$ in it, $\delta(z)/\delta(0)$
will be $k$-independent, too. We will also suppose that $c_s$ is
not too small, so that $k\gg a/\lambda_J$ for all scales of
interest, in particular, $c_s\gg 0.01$ if we consider scales up to
$100(1+z)^{-1}$ ~Mpc.

The dimensionless physical distance
\beq\label{eq:E}
E = a(t_0)H_0\int_{t}^{t_0} \frac{dt}{a(t)} = H_0\int_0^z 
\frac{dz_1}{H(z_1)}~,
\eeq
where $t_0$ is the present moment, plays a key role in measurements of
the background universe using standard rulers and candles.  $E$ is
proportional to the conformal time measured from the present to a
moment in the past.  It is related to the luminosity distance, $d_L$,
via
\ber\label{Edl1}
\frac{H_0d_L(z)}{1+z} &=& 
\frac{1}{\sqrt{\vert\Omega_{K}\vert}}\sin\lbrace \sqrt{\vert\Omega_{K}\vert}E(z)\rbrace ~,
 ~~ \Omega_K<0\\
\label{Edl2}
\frac{H_0d_L(z)}{1+z} &=&E(z)~, \hskip 3.8cm 	\Omega_K=0\\
\label{Edl3}
\frac{H_0d_L(z)}{1+z} &=&
\frac{1}{\sqrt{\Omega_{K}}}\sinh\lbrace \sqrt{\Omega_{K}}E(z)\rbrace~, \hskip0.6cm \Omega_K>0
\eer
%\beq \label{E-rel}
%E(z) = \frac{H_0d_L(z)}{1+z} = H_0(1+z)d_A(z)~,
%\eeq
where $\Omega_K \equiv 1-\Omega_{\rm total}$.  The following
relationship between the luminosity distance $d_L$ and the angular
size distance $d_A$ holds in a metric theory of gravity: $d_L =
(1+z)^2d_A$.  Rewriting Eq~(\ref{eq:diff}) in terms of Eq~(\ref{eq:E})
and using the fact that $\rho_m\propto (1+z)^3$ , we obtain:
\beq \label{eq:diff1}
\left(\frac{\ddel}{1+z(E)}\right)^{\prime}=\frac{3}{2}\Omega_{0m}
\delta~,
\eeq
where the prime denotes a derivative with respect to $E$. It is
straightforward to transform Eq~(\ref{eq:diff1}) into the following
set of integral equations for $\del (E)$ and its first derivative
\citep{ss06}:
\ber\label{eq:int}
\del (E) &=& 1+\dd \int_0^E [1+z(E_1)] dE_1 + \frac{3}{2} \om 
\int_0^E [1+z(E_1)] \left( \int_0^{E_1} \del (E_2) dE_2 \right) dE_1 \\
\label{eq:int1}
\ddel (E) &=&  \dd [1+z(E)] +  \frac{3}{2} \om [1+z(E)] \int_0^{E} 
\del (E_1) dE_1 \,\,,
\eer
where $\del$ is normalized to $\del_0\equiv\del(z=0) = 1$. Note the
remarkable fact that, in contrast to formulas used in the
reconstruction of $H(z)$ from $d_L(z)$ \citep{star98,DE_recon} or
$\delta(z)$ \citep{star98} which require taking a derivative of
observational data with respect to the redshift, this formula contains
integrations of observational data only, which is a sound operation
for noisy data. %Eqs~(\ref{eq:diff}, \ref{eq:diff1}-\ref{eq:int1})
%remain the same in the spatially curved case, the only equation in
%this reconstruction method which does change is Eq~(\ref{E-rel}).

By solving the above equations we can calculate the linear {\em growth
  factor}
\beq
g(z)\equiv (1+z)\del(z)~,
\eeq 
which represents the ratio of $\del(z)$ in the presence of dark energy
to that in SCDM without a cosmological constant. Another quantity of
interest is the {\em growth rate}
\beq
f(z) = \frac{d\ln\del}{d\ln a} = - \frac{1+z}{H(z)} \ \frac{\ddel (E)}{\del(E)} \,\,.
\eeq

To solve Eq~(\ref{eq:int}) we start with initial guess values for
$\del (E)$ and $\ddel (E)$ and iteratively solve for $\del (E)$,
calculating $\ddel (E)$ in the successive iterations as the difference
between adjacent values of $\del (E)$, \ie $\ddel_i = \Delta \delta_i
/\Delta E_i$. For calculating $f(z)$ we require to estimate $H(z)$ as
well. We obtain this quantity by differentiating the noisy data $E(z)$
using a finite differencing method . This naturally amplifies the
noise in the final results, so we expect the results for $f(z)$ to be
slightly noisier. However, typically the difference in $f(z)$ between
two models of dark energy is greater than the difference in $g(z)$, so
despite the greater noise, we expect $f(z)$ to be useful for
discriminating dark energy models. This method does not require prior
knowledge of the parameter $\dd$, is robust to changes in the initial
guess values and gives exact results for $g(E)$ and $f(E)$ for
noiseless data. For data with errors, naturally the result is noisier,
however, as we will show in the succeeding sections, we will be able
to put reasonable constraints on $g$ and $f$ using this method.

Data noise can also be decreased using smoothing techniques. In what
follows we shall use the lognormal smoothing scheme proposed in
~\citep{smooth} which has been shown to be reasonably unbiased and
efficient. It constructs a smooth quantity, $E^s$, from a noisy one,
$E(z_i)$, via the ansatz ~\citep{smooth} 
\beq\label{eq:smooth}
E^s(z) = \sum_i E(z_i) {\rm exp} \left( \frac{-{\rm ln}^2 
\frac{1+z_i}{1+z}}{2 \Delta^2} \right) / \sum_i {\rm exp} 
\left( \frac{-{\rm ln}^2 \frac{1+z_i}{1+z}}{2 \Delta^2} \right) \,\,, 
\eeq
where $\Delta$ is the smoothing scale \citep[see also][]{shaf07}. We
take $\Delta \simeq 1/N$ where $N$ is the total number of
observations. Choosing this small value of $\Delta$ leaves the results
unbiased.

From the manner in which Eqs~(\ref{eq:diff1}-\ref{eq:int1}) have been
obtained, reasons as to why the linearized growth function
$\delta_{\rm obs}(z)$ determined from actual observations of large
scale structure may differ from $\del(z)$ reconstructed using our
method follow immediately:

\begin{enumerate}

\item 
$\rho_m$ is not proportional to $(1+z)^3$. This happens even in GR if
the energy-momentum tensor of physical DE is not covariantly conserved
separately, either due to the existence of a direct non-gravitational
interaction between DE and DM \citep[see][and references
  therein]{dm_de,dmde1,dmde2,dm_de1}, or because DE and DM constitute
some unique entity as occurs in unified DM--DE models such as the
Chaplygin gas \citep{Kam01} and its generalizations.  In such models
DE is partially clustered with DM on small scales, and this can result
in the appearance of significant $k$-dependent terms in the equation
for $\del$, so that the growth factor $g(z)$ becomes $k$-dependent. In
particular, the latter effect is especially crucial for the
generalized Chaplygin gas model, \citep[see the recent paper][and
  references to previous papers therein]{Gor08}.

\item 
GR is modified, DE is geometrical. Then $G$ in Eq~(\ref{eq:diff})
becomes some effective quantity $G_{eff}$ which may be both time and
scale dependent. A noticeable value of $\phi-\psi$ may also arise even
in the absence of free-streaming particles. However on small scales
this value is strongly restricted by Solar system tests of gravity
(which don't suggest any such effect). Geometrical models of DE, which
include Braneworld models and models using scalar-tensor and $f(R)$
gravity, have more degrees of freedom than GR, so it is natural that
in this case, the linearized perturbation equation for $\del$ shows a
departure from the Newtonian form, Eq~(\ref{eq:diff}). For instance in
extra dimensional scenario's \citep{dgp,brane}, the presence of the
fifth dimension (the bulk) can influence the behaviour of
perturbations residing on the brane
\citep{f_dgp,brane_pert_DGP,brane_pert_DGP2,brane_pert} making them
significantly $k$-dependent even on scales much smaller than the
Hubble distance \citep{brane_pert}.  The same effect arises in viable
DE models in $f(R)$ gravity \citep{fR,fR_star,fR_tsuji} \citep[see the
  recent review][for numerous papers on $f(R)$ gravity as a
  whole]{SF08}. On the other hand, in some cases $G_{eff}$ and $g(z)$
may remain scale-independent even on small scales, though they acquire
a non-trivial, non-GR, time dependence. This occurs, for instance, in
the Dvali-Gabadadze-Porrati (DGP) extra-dimensional model considered
below as well as in scalar-tensor DE models with a large current value
of the Brans-Dicke parameter $\omega_{BD}$ \citep{beps00, gp08}.

\end{enumerate}

Thus, a comparison of the observed and reconstructed density contrast
could help shed light on the nature of dark energy. While it is
encouraging that future observations \citep{ska,guzo,percival,euclid}
of large scale structure may make possible the determination of
$\delta_{\rm obs}(z)$, in this paper we focus on reconstructing $\del$
using observations of high redshift type Ia supernovae and baryon
acoustic oscillations.

\subsection{Data used}

The method outlined in the previous section would be applicable to any
observation which contains a measurement of $E(z)$, \eg measurements
of luminosity distance or angular diameter distance. We shall use real
data and mock data based on simulations of supernova type Ia data and
the angular diameter distance from baryon acoustic oscillations, to
test this method.

\n{\bf Supernova Data :}

\smallskip
The lightcurves of Type Ia supernovae show them to be ``calibrated
candles'', therefore they are of enormous significance in cosmology
today. The luminosity distance of Type Ia SNe provide us with a direct
measurement of the acceleration of the universe, thus leading to
constraints on the dark energy parameters. SNe data is in the form
$\lbrace m_B, z, \sigma_{m_B}, \sigma_z \rbrace$, where the magnitude
$m_B$ is related to $d_L(z)$ as
\beq\label{eq:sne}
m_B = 5 {\rm log_{10}}[H_0d_L(z)] + {\cal M}\,\,,
\eeq
${\cal M}$ being a noise parameter usually marginalized over.

It should be noted that supernova data, on its own, is unable to break
the degeneracy between dark energy and spatial curvature.  The CMB, on
the other hand, places stringent constraints on $\Omega_K $ and
strongly suggests that the universe is spatially flat, in agreement
with predictions made by the inflationary scenario. In this paper we
shall work under the assumption that $\Omega_K=0$ and use (\ref{Edl2})
to relate $d_L \to E(z)$, with the latter playing the key role in our
reconstruction exercise (\ref{eq:int}).

Currently there are around $300$ published SNe with the furthest
observed one at a redshift of $z = 1.7$ \citep{union}, and average
error of $\sigma_{m_B} \simeq 0.15$. Future space-based projects such
as the Joint Dark Energy Mission (JDEM) \citep{jdem} are expected to
observe about $2000$ SNe with errors of $\sigma_{m_B} = 0.07$. To
date, SNe are the most direct evidence for dark energy, and in this
paper we shall primarily use SNe data to constrain the growth
parameters for different dark energy models.

\n{\bf BAO data}

\smallskip
At present, baryon acoustic oscillations are believed to be the method
least plagued by systematic uncertainties, therefore the detection of
the first baryon acoustic oscillation scale \citep{bao} has led to the
speculation that BAO may in future become a potent discriminator for
dark energy. Standing sound waves that propagate in the opaque early
universe imprint a characteristic scale in the clustering of matter,
providing a ``standard ruler''. Since the sound horizon is tightly
constrained by cosmic microwave background (CMB) observations,
measuring the angle subtended by this scale determines a distance to
that redshift and constrains the expansion rate. The radial and
transverse scales give measurements of $[r_s H(z)]/c$ and $r_s/[(1+z)
  d_A(z)]$ respectively, where $r_s$ is the sound horizon obtained
from CMB. These quantities are correlated, and the present BAO data is
not sensitive enough to measure both quantities independently \citep[see
however the recent papers][]{ben08, gaz08}, but future surveys are
expected to give independent measurements of $d_A(z)$ and $H(z)$
\citep{bao1}. Future BAO surveys such as BOSS \citep{sdss} should
therefore place tighter constraints on dark energy parameters.

\section{Results}\label{res}

We first use Supernova data to reconstruct the growth parameters. We
simulate data according to two theoretical models :
\begin{itemize}
\item
Model 1 : A cosmological constant model with $w=-1, \om=0.27, H_0=72 \ 
{\rm km/s/Mpc}$.
\item
Model 2: A variable dark energy model with the equation of state given by
\beq\label{eq:model2}
w(z) = w_0+\frac{w_az}{1+z}~, \ w_0 = -0.9, ~w_a = 0.3 \,\,,
\eeq
and with the same values of $\om, H_0$ as Model 1. Note that the Models 
1 and 2 provide excellent agreement with the current CMB+BAO+SNe data
\citep{wmap05}. The Model 2 has $w>-1$ everywhere, so it can be realized
by quintessence with some potential for which Eq~(\ref{eq:diff}) is valid.

\end{itemize}

Two different data distributions are used, set A resembles the quality
of data available at present, and set B is modeled on expected future
surveys.
\begin{itemize}
\item
Set A : $\sim 300$ SNe, with the redshift distribution and errors of
the Union dataset \citep{union}. For this dataset, on average,
$\sigma_{m_B} \simeq 0.15$, but a few SNe have very high errors of the
order of unity. Since the method of integration would not work very
well for very noisy data, and a single datapoint with large noise
would affect the results of all datapoints after it, we restrict the
analysis to SNe with $\sigma_{m_B} < 0.7$. By rejecting only $10$
datapoints with this criterion, we enhance the results by a
significant amount.

\item
Set B : $\sim 2000$ SNe, with the redshift distribution and errors
($\sigma_{m_B} \sim 0.07$) expected from future surveys such as the
JDEM \citep{snap}. The data covers a redshift range of $z = 0.1-1.7$
with a larger concentration of supernovae in the midrange redshifts
($z = 0.4-1.1$). The errors considered here are statistical only, we
do not consider systematic errors, which are expected to be better
controlled in the future with larger datasets.
\end{itemize}
For both cases, we marginalize over $\om=0.27 \pm 0.03$. Supernova
data is unable to break the degeneracy between dark energy and
curvature of the universe. In order to measure the growth parameters,
we therefore consider only a flat universe, which is the preferred
model from current CMB observations.

\begin{figure*} 
\centering
\begin{center}
\epsfxsize=6.8in
\epsffile{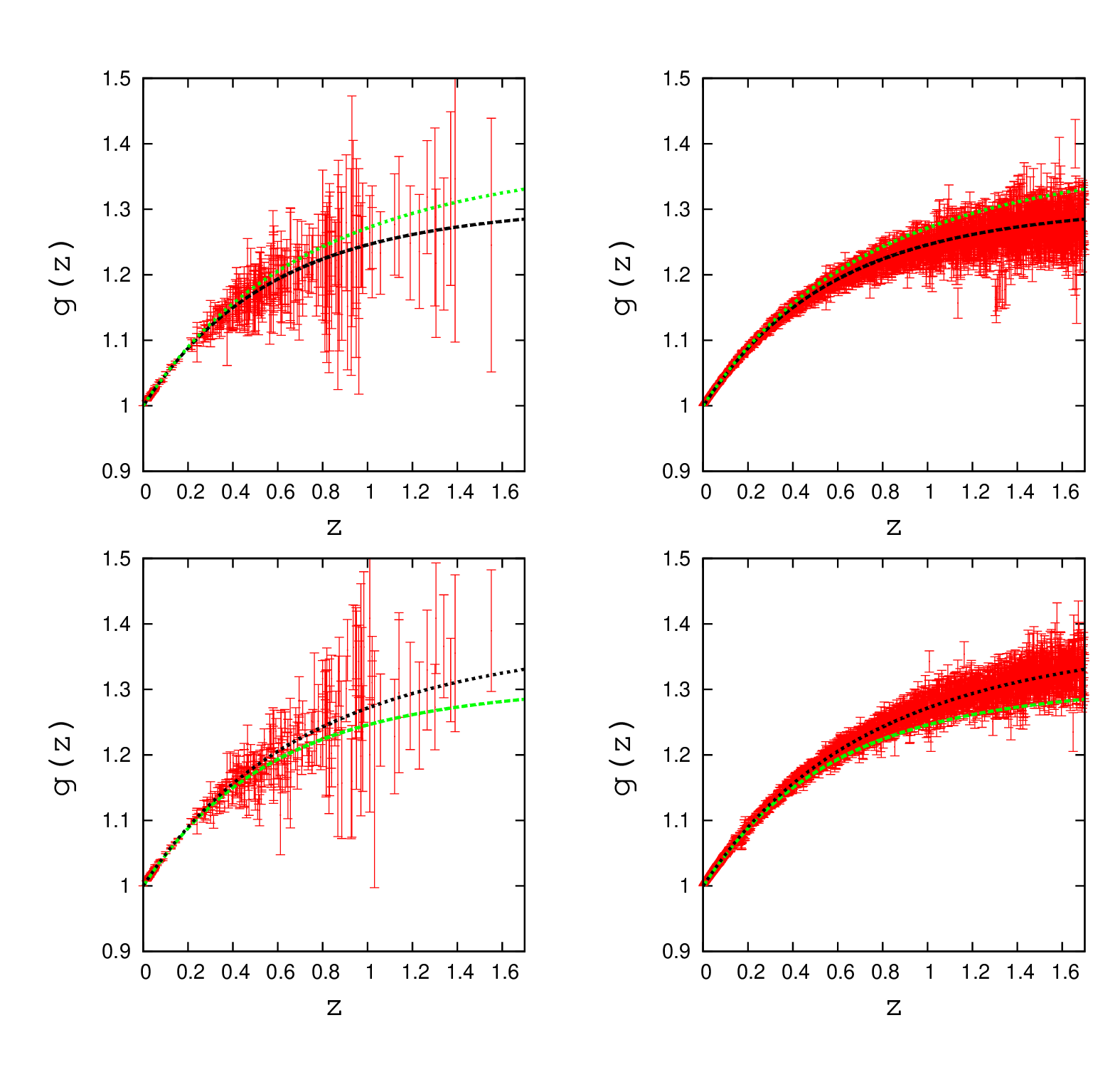}
\end{center}
\vspace{0.0cm}
\caption{\small 
Reconstructed linear growth factor $g(z)$ for different datasets. The
top panels show the results for Model 1 ($\ld$CDM) using
Union-like (set A, left panel) and JDEM-like (set B, right panel) SNe
datasets, while the bottom panels show results for Model 2 (variable
$w$, eq~(\ref{eq:model2})) using set A (left panel) and set B (right
panel). In each figure, the black dotted line represents the true
model, while the green dashed line represents the other model. The red
solid lines show the $1\sigma$ error bars for the integral
reconstruction using eq~(\ref{eq:int}).  }
\label{fig:g}
\end{figure*}

Fig~\ref{fig:g} shows the results for the linear growth factor $g(z)$
for both datasets and for the two different cosmological models. We
see that for both models, set A results in rather noisy reconstruction
(left panel), since the errors on the SNe are quite high. This is
especially true at high redshifts ($z > 0.7$) where the sparse
sampling affects the integral reconstruction scheme adversely. At
$z=0.3, \ g(z)$ is constrained to $\sim 5 \%$. For JDEM-like data (set
B) however, $g(z)$ is reconstructed more accurately, and has low
errors at low redshifts (right panel). At $z = 0.3$, $g(z)$ is
constrained accurately to $\sim 2 \%$ for both models for set B, while
at $z=1$, $g(z)$ is constrained to $\sim 4\%$.

Fig~\ref{fig:f} shows the reconstruction of the growth rate $f(z)$. As
before, the results for set A are poor, with $f(z)$ constrained to
$\sim 6 \%$ at $z = 0.3$. The results for Set B are reasonable,
however, the errors are slightly larger in this case than for $g(z)$,
since there is an additional error from the calculation of $H(z)$ from
$E(z)$.  At $z = 0.3$, $f(z)$ is constrained accurately to $\sim 3 \%$
for both models for set B, while at $z=1$, $f(z)$ is constrained to
$\sim 8\%$. We also note that the quantity $f(z)$ has slightly greater
discriminatory power than $g(z)$, since typically the growth factor
shows rather less variation between different dark energy models as
compared to the growth rate at any given redshift. Therefore, even
though $f(z)$ is slightly noisier, for set B, Model 1 and Model 2 can
be discriminated at $1\sigma$ using $f(z)$.

\begin{figure*} 
\centering
\begin{center}
\epsfxsize=6.8in
\epsffile{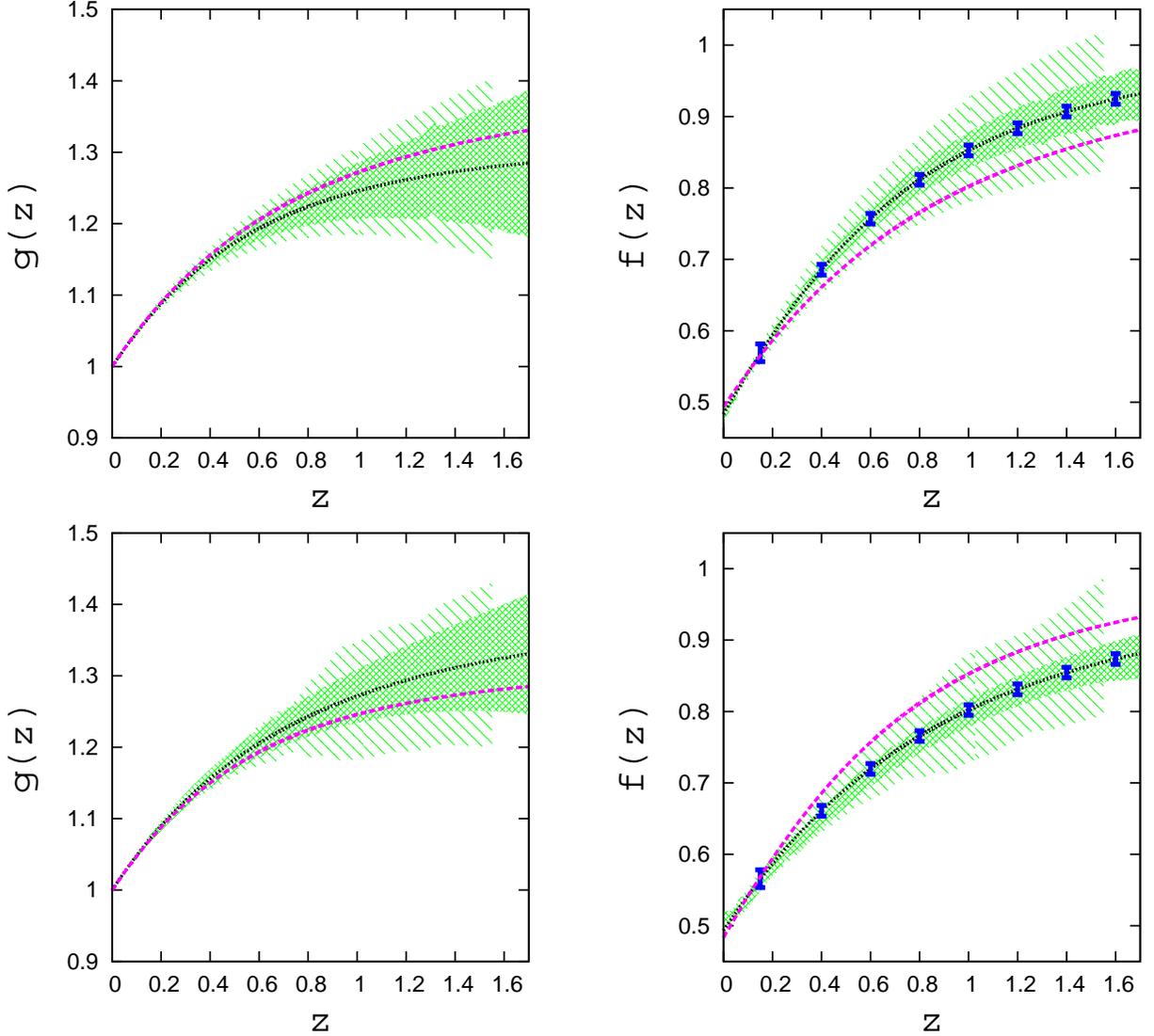}
\end{center}
\vspace{-0.6cm}
\caption{\small 
Reconstructed growth parameters for different datasets using the
smoothing scheme Eq~(\ref{eq:smooth}) on the integral reconstruction
method, eq~(\ref{eq:int}). The top panels show the results for Model 1
($\ld$CDM) for the growth factor $g(z)$ (left panel) and the
growth rate $f(z)$ (right panel). The bottom panels show the results
for Model 2 (variable $w$, eq~(\ref{eq:model2})) for $g(z)$ (left
panel) and $f(z)$ (right panel). In each figure, the black dotted line
represents the true model, while the pink dashed line represents the other model. The green dashed shaded area represents the
$1\sigma$ errors for the integral reconstruction of set A
(Union-like), while the green hatched shaded area represents the
reconstruction for set B (JDEM-like). The blue vertical lines in the
right panel show the expected observational constraints from Euclid
\citep{euclid}.  }
\label{fig:smooth}
\end{figure*}

If the data is first smoothed with the smoothing scheme,
Eq~(\ref{eq:smooth}), the results improve, especially for set A which
has much noisier data, as seen in figure~\ref{fig:smooth}. The results
for $f(z)$ improve markedly for both datasets. This is because an
additional quantity $H(z)$ is required for obtaining $f(z)$, and a
smoother $E(z)$ leads to a much more accurate estimation of
$H(z)$. Errors on $g(z)$ and $f(z)$ are $\sim 1 \%$ and $\sim 1.5 \%$
respectively at $z=0.3$, and $\sim 3 \%$ and $\sim 6 \%$ respectively
at $z=1$ for Model 1 with JDEM like data. Model 2 gives similar
constraints. The results for the growth parameters are summarized in
Table~\ref{tab:lcdm} for Model 1, and in Table~\ref{tab:varw} for
Model 2. We see that this method obtains quite reasonable constraints
on the growth parameters at low redshifts for the set B, therefore it
can be used successfully to constrain growth parameters from future
SNe data. It should be noted that, for future SNe data to accurately
constrain the growth parameters, it is important to keep the SNe
systematics under control ($\sigma_{sys} \lleq 0.05$). A systematic
error of $\sigma_{sys} = 0.1$ (as on the current data) would weaken
all constraints significantly.

\subsection{Dependence on nature of data}

\begin{figure*} 
\centering
\begin{center}
\epsfxsize=6.8in
\epsffile{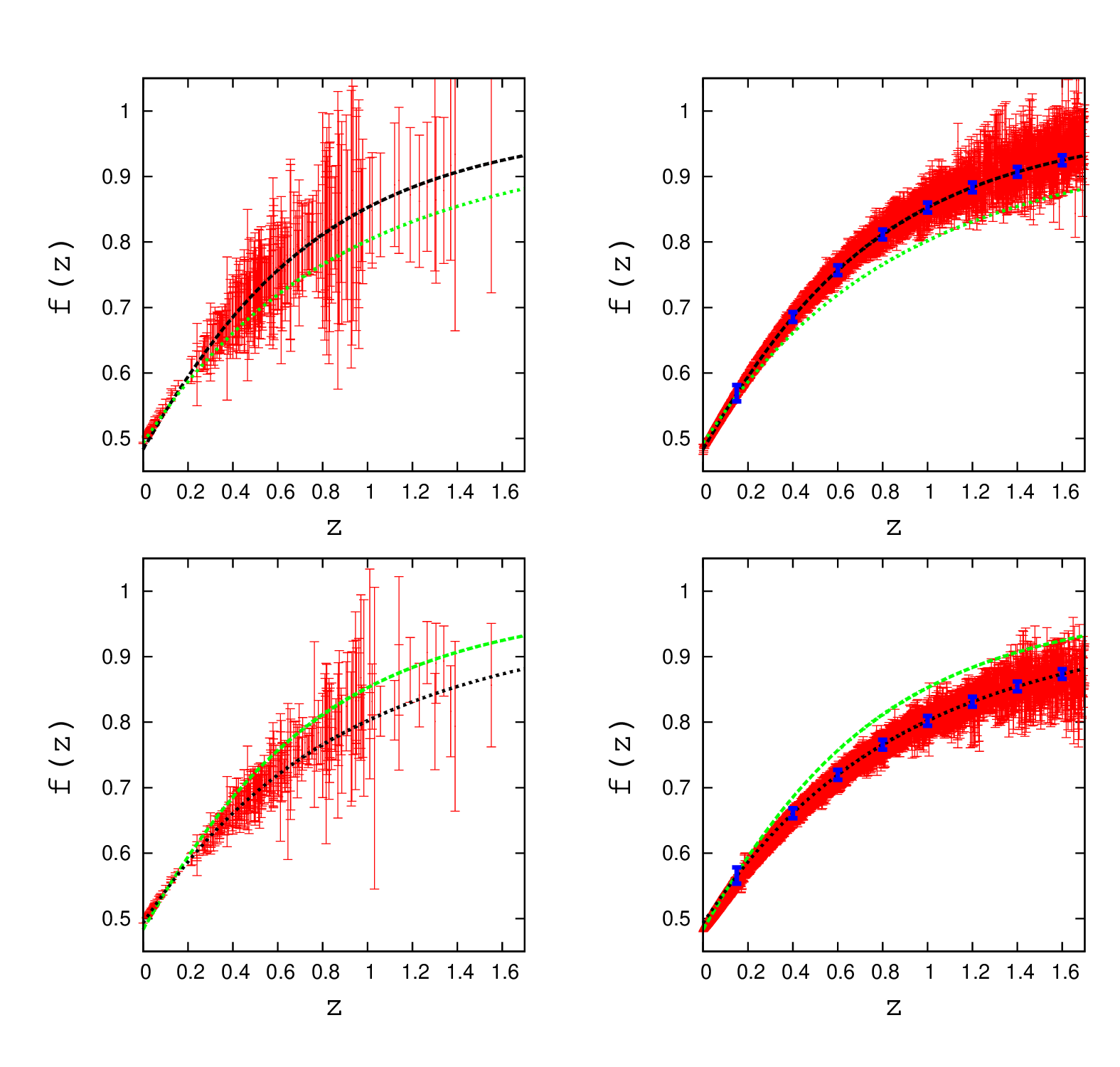}
\end{center}
\vspace{0.0cm}
\caption{\small 
Reconstructed growth rate $f(z)$ for different datasets. The top
panels show the results for Model 1 ($\ld$CDM) using Union-like
(set A, left panel) and JDEM-like (set B, right panel) SNe datasets,
while the bottom panels show results for Model 2 (variable $w$,
eq~(\ref{eq:model2})) using set A (left panel) and set B (right
panel). In each figure, the thick black dotted line represents the
true model, while the green dashed line represents the other
model. The red solid lines show the $1\sigma$ error bars for the
integral reconstruction using eq~(\ref{eq:int}) The blue vertical
lines in the right panel show the expected observational constraints
from Euclid \citep{euclid}.  }
\label{fig:f}
\end{figure*}

\begin{figure*} 
\hspace{-1.3cm}
\epsfxsize=8.6in
\epsffile{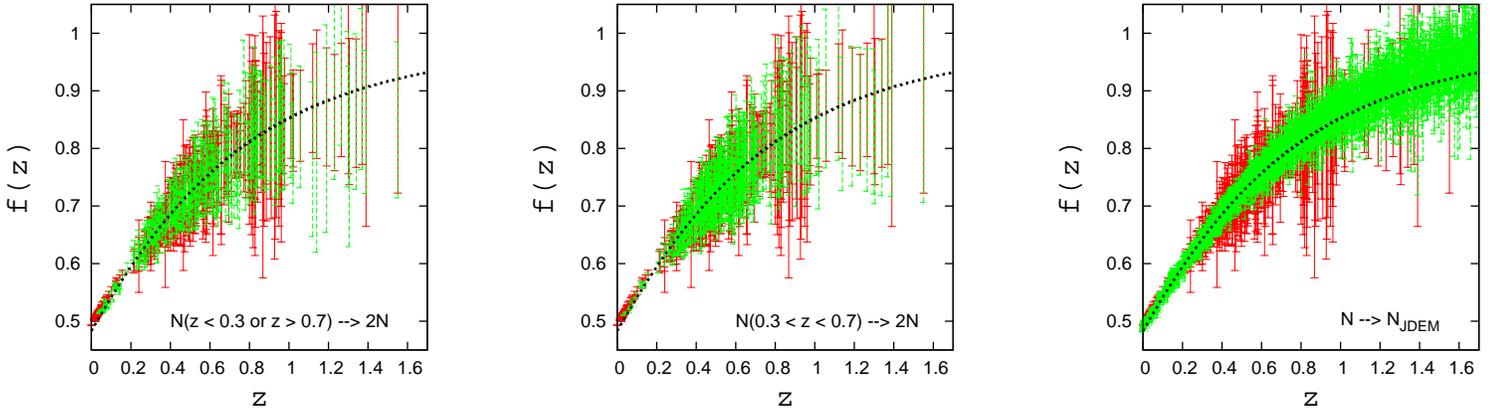}
\vspace{0.0cm}
\caption{\small 
Reconstructed growth rate $f(z)$ for model 1 ($\ld$CDM) using
various redshift distributions. We use (a) set A (Union-like) with
number of SNe doubled at low and high redshifts (left panel) (b) set A
with number of supernova doubled for mid-range SNe (center panel) and
(c) JDEM-like (set B) redshift distribution with Union-like (set A)
errors (right panel). In each panel, the red solid lines depict
$1\sigma$ error bars on set A, while the green dashed lines show the
$1\sigma$ error bars on set A modified according to (a), (b), (c). The
black dotted line represents the true model.  }
\label{fig:N}
\end{figure*}

We now check how the results change if the redshift distribution or
error distribution is changed. To study the dependency on the number
of SNe, we use three redshift distributions-- (a) set A ($\sim 300$
SNe) with double the number of supernovae at low ($z < 0.3$) and
high($z>0.7$) redshifts, (b) set A with double the SNe at mid-range
($0.3 < z < 0.7$) redshifts, and (c) a distribution with the JDEM (set
B) redshift distribution ($\sim 2000$ SNe) with errors of the order of
the Union (set A) SNe. The results for Model 1 are shown in
figure~\ref{fig:N}. We see that doubling the number of SNe in a
particular redshift bin changes the results very slightly. This is to
be expected because when integrating noisy data, having a larger
number of points with the same amount of noise does not improve
results significantly. Increasing the total number of SNe by a
significant amount (nearly seven times, as in right panel) does
improve the scatter, but the results still do not compare with those
of set B (fig~\ref{fig:f}, top right panel) which has the same number
of supernovae but smaller errors.

\begin{figure*} 
\hspace{-1.3cm}
\epsfxsize=8.6in
\epsffile{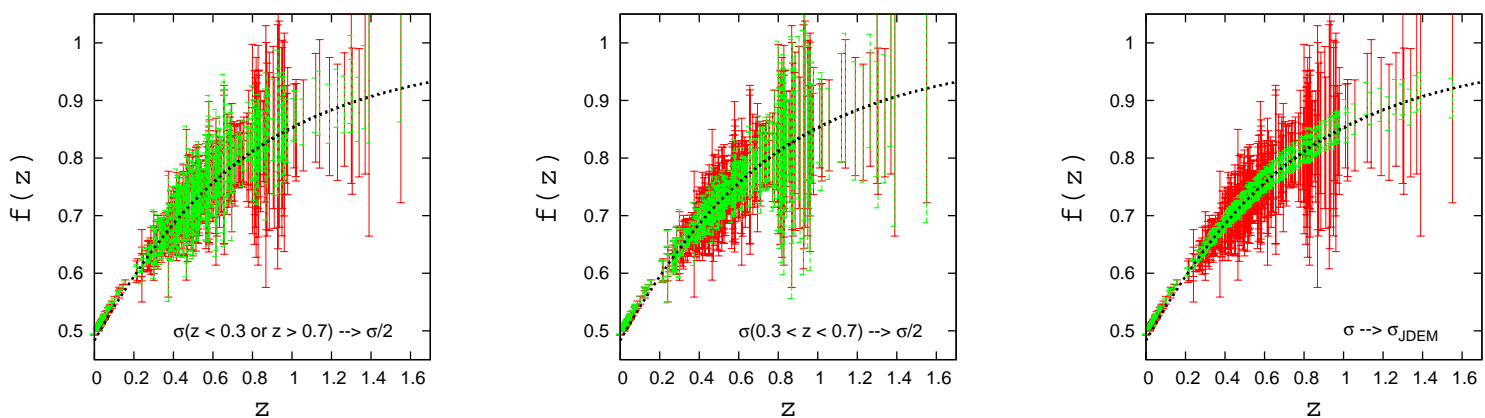}
\vspace{0.0cm}
\caption{\small 
Reconstructed growth rate $f(z)$ for model 1 ($\ld$CDM) using
various error distributions. We use (a) set A (Union-like) with errors
halved at low and high redshifts (left panel) (b) set A with errors
halved for mid-range SNe (center panel) and (c) set A with JDEM-like
errors for each SNe (right panel).  In each panel, the red solid lines
depict $1\sigma$ error bars on set A, while the green dashed lines
show the $1\sigma$ error bars on set A modified according to (a), (b),
(c). The black dotted line represents the true model.  }
\label{fig:err}
\end{figure*}

We now study the effect of the errors. Once again we study three
distributions --- (a) set A with the errors halved for $z<0.3$ and
$z>0.7$ redshift bins, (b) set A with errors halved in the $0.3<z<0.7$
redshift bin, and (c) set A with errors replaced by JDEM-like errors
on all SNe. The results for Model 1 are shown in
figure~\ref{fig:err}. We see that in this case, decreasing the errors
at low redshift or high redshift changes the results very
slightly. This is because there are very few points at low redshift so
they do not affect the integration process strongly, while the high
redshift points cannot affect the low redshift points. The results in
the redshift range $0.3 < z < 0.7$ become better if the mid-range SNe
have lower errors. As we see in the right panel of fig~\ref{fig:err},
decreasing the errors to JDEM errors gives results almost identical to
the results for Set B (fig~\ref{fig:f}, top right panel) , even though
the number of points is much less for set A. Thus we find that this
method would work quite well even for a reasonable number of
supernovae (of the order of a few hundred) provided the errors were
tightly constrained.

Since the high errors of set A make it unsuitable for this
reconstruction approach, in the next sections we will use the set B to
study the robustness of the results to various other factors.

\subsection{Growth rate from $w(z)$}

We may also calculate the growth rate $f$ from the supernova data via
the equation of state using the following approximation \citep{growth}
:
\ber\label{eq:f_w}
f(z) &\simeq& \Omega_m(z)^\gamma = \left[ \frac{\om (1+z)^3}{H^2(z)} \right]^{\gamma}\\
\gamma(z) &=& \frac{3}{5-\frac{w}{1-w}}+\frac{3}{125}\frac{(1-w)
(1-\frac{3}{2}w)}{(1-\frac{6}{5}w)^3}\left(1-\Omega_m(z)\right) + 
{\cal O}\lbrack\left(1-\Omega_m(z)\right)\rbrack^2 \,\,,
\eer
where the equation of state $w(z)$ may be calculated using a
likelihood parameter estimation from the luminosity distance. This
approximation works quite well for a large number of physical dark
energy models with a constant or slowly changing $w$ including LCDM, 
for which $\gamma \simeq 0.55$
\citep{growth,ff}.  We use the familiar CPL fit \citep{cpl, lin-cpl} :
\ber\label{eq:w}
w(z) &=& w_0+\frac{w_az}{1+z}~,\\
H^2(z) &=& H_0^2 \left[ \om (1+z)^3 +(1-\om) (1+z)^{3(1+w_0+w_a)} 
e^{3 w_a (1/(1+z)-1)} \right] \,\,.
\eer
A likelihood parameter estimation is expected to lead to smaller
errors, but the drawback of this method is that the result may be
biased due to the parameterization. Also the errors on $w(z)$ would
propagate extremely non-linearly to $f$ and therefore the result for
$f(z)$ would be much less trustworthy.

\begin{figure*} 
\centering
\begin{center}
\epsfxsize=6.8in
\epsffile{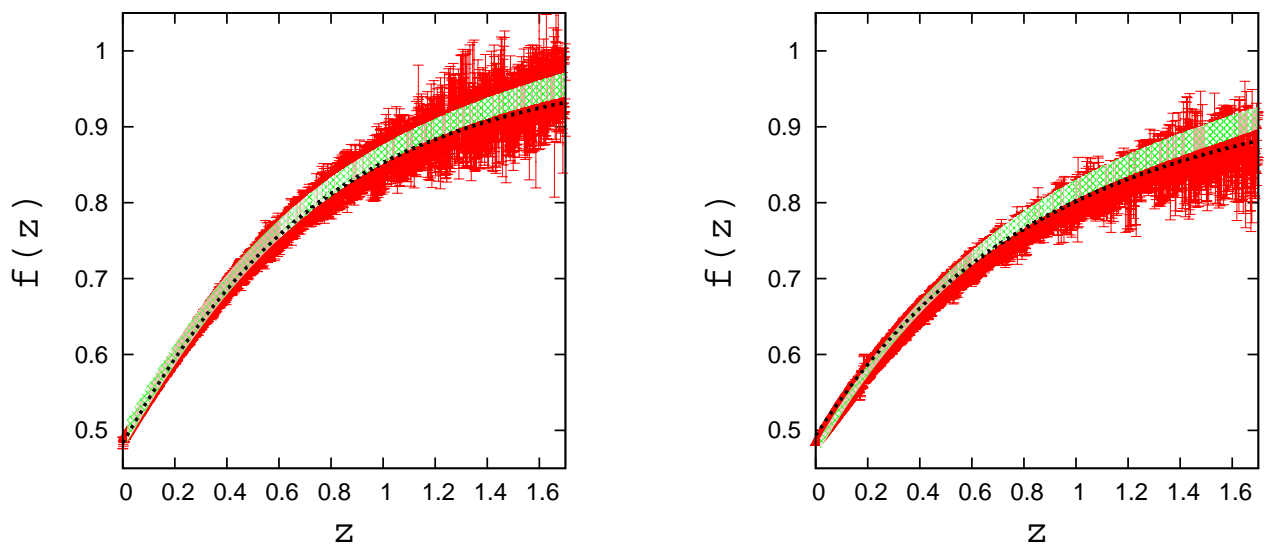}
\end{center}
\vspace{0.0cm}
\caption{\small 
Reconstructed growth rate $f(z)$ for model 1 (left panel) and model 2
(right panel) using set B (JDEM-like) with different reconstruction
methods. The red solid lines show the $1\sigma$ limits for
reconstructed $f(z)$ using the integral reconstruction method,
eq~(\ref{eq:int}), while the green hatched region shows the $1\sigma$
limits for $f(z)$ using $w$ parameterization,
eqs~(\ref{eq:f_w}),~(\ref{eq:w}). The black dotted line represents the
true model.  }
\label{fig:f_w}
\end{figure*}

Figure~\ref{fig:f_w} shows the reconstructed $f(z)$ for Model 1 and 2
for set B. As expected, the errors are lower that those for our
reconstruction method. However, it is also noteworthy that the
resulting confidence levels are {\em not symmetric} around the true
value, in fact at higher redshifts, the true model appears to be on
the verge of being ruled out ! These results are commensurate with
those found in \citep{hu}, where reconstruction of the growth
parameters through $w$ leads to biases in the growth parameter results
even though $w$ is recovered accurately. This is due to the fact that
errors propagate non-linearly from $w$ to $f(z)$. We therefore
conclude that, when reconstructing the growth parameters from
supernova data, it is better to reconstruct the quantities directly,
rather than reconstructing them indirectly from the energy density or
equation of state.

\subsection{Dependence on $\om$}

\begin{figure*} 
\centering
\begin{center}
\epsfxsize=6.8in
\epsffile{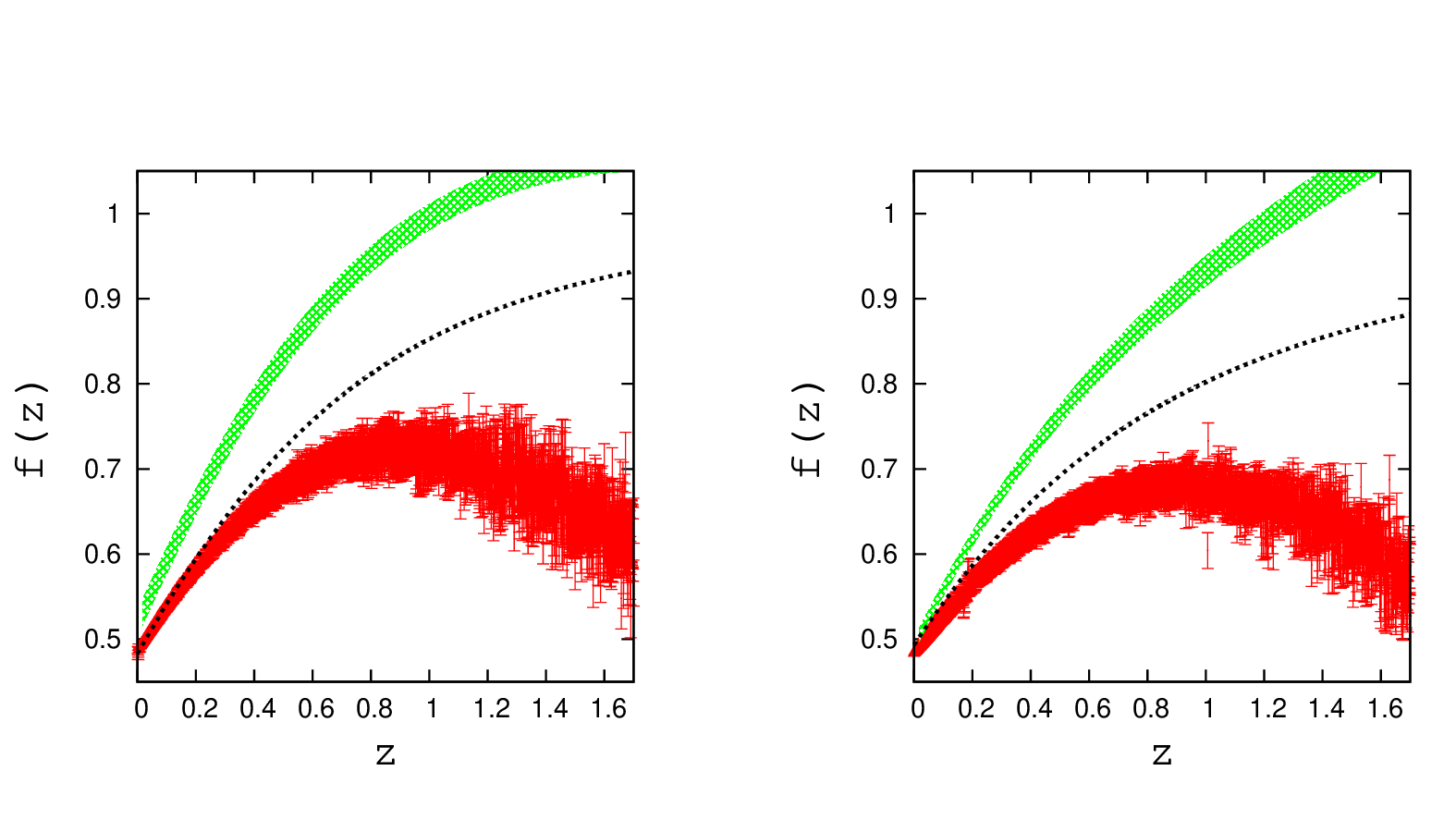}
\end{center}
\vspace{0.0cm}
\caption{\small 
Reconstructed growth rate $f(z)$ for model 1 (left panel) and model 2
(right panel) for set B (JDEM-like) with different reconstruction
methods, using $\om=\om({\rm true}) + 0.03$. The red solid lines show
the $1\sigma$ limits for reconstructed $f(z)$ using the integral
reconstruction method, eq~(\ref{eq:int}), while the green hatched
region shows the $1\sigma$ limits for $f(z)$ using $w$
parameterization, eqs~(\ref{eq:f_w}),~(\ref{eq:w}). The black dotted
line represents the true model. Note that the results for the two
reconstructions lie on opposite sides of the true value of $f(z)$.  }
\label{fig:om}
\end{figure*}

Supernova data do not simultaneously constrain information on $\om$
and dark energy parameters. To reconstruct dark energy parameters, it
is necessary to place constraints on $\om$ from other observations. In
the calculations so far, we have marginalized over the true fiducial
value for $\om$. However, since there is considerable uncertainty as
to the real value of the matter density, we check how using {\em
incorrect} values of $\om$ may bias our analysis. (It is well known
that an incorrect value of $\om$ can significantly bias the results
for DE \citep{smooth,sss08}.) The fiducial universe for model 1
contains $\om=0.27$. We now choose a different, incorrect, value of
$\om=0.3$ for marginalization and proceed to analyze the data using
both the integral reconstruction method and the likelihood parameter
estimation of $w$ outlined in the previous section. The results are
shown in figure~\ref{fig:om}. We see that choosing a higher value of
$\om$ gives biased results in both methods, but interestingly enough,
the biases are in {\em opposite directions} ! In case of the integral
reconstruction method, a higher value of $\om$ leads to a lower value
of $f(z)$ at high redshifts, whereas for the $w$ parameterization, a
higher value of $\om$ leads to a higher value of $f(z)$.

These results may be understood as follows. For the reconstruction
from $w$, we see from eq~(\ref{eq:f_w}) that $f(z)$ changes primarily
due to the change in the matter density $\om (1+z)^3$, since the value
of $\gamma$ does not vary very strongly with $w$, and $H^2(z)$ is
constrained by the data. Choosing a higher value of $\om$ would result
in the choice of a different $w(z)$ which would lead to nearly the
same $H(z)$ as that for a lower value of $\om$, and $\gamma$ would
also not change by much. However, the quantity $\om (1+z)^3$ would
increase proportionate to $\om$. Therefore a higher value of $\om$
would simply result in a higher value of $f(z)$. In the case of the
integral reconstruction however, we see from eq~(\ref{eq:int}) that
both $\del$ and $\ddel$ depend on $\om$. In $\del$ the leading term is
unity and the other two terms containing $\dd$ and $\om$ are at about
an order of magnitude smaller. In $\ddel$ the two terms containing
$\dd$ and $\om$ are of the same order and opposite sign. The $\om$
term contributes by making $\ddel$ less negative. Therefore increasing
$\om$ increases $\del$ slightly and decreases the absolute value of
$\ddel$ by a larger amount, so that the ratio between $\del$ and
$\ddel$ becomes a smaller negative quantity. Since $f(z)$ is
essentially this ratio, this means that $f(z)$ also decreases with
increasing $\om$. Therefore, choosing a wrong value of $\om$ causes
the two different methods of reconstruction to be biased in opposite
directions. This leads to the interesting conclusion that, provided
other systematics are under control, comparing the integral
reconstruction method with the standard likelihood estimation would
give us a valuable consistency check on the accuracy of the prior
chosen for $\om$.

\subsection{Reconstruction for a toy modified gravity model} \label{dgp}

An influential Braneworld model was suggested by
Dvali-Gabadadze-Porrati \citep{dgp}. The expansion history for this
model is given by
\beq\label{eq:dgp}
H(z) = H_0 \left[ \left(\frac{1-\om}{2}\right)+\sqrt{\om (1+z)^3+
\left(\frac{1-\om}{2}\right)^2} \right]\,\,.
\eeq

For physical models of dark energy, the growth rate is well
approximated by eq~(\ref{eq:f_w}), for instance $\gamma \simeq 0.55$
for LCDM \citep{growth,ff,ff1,ff2,ff3}. This equation is not valid
however if the observed acceleration originates from a modification of
the equations of general theory of relativity; in the DGP Braneworld
theory, the growth rate is approximated by \citep{f_dgp}
\beq\label{eq:f_dgp}
f (z) \simeq \Omega_m(z)^{0.68} \,\,.
\eeq
This is the growth rate which would be measured through galaxy
redshift distortions or weak gravitational lensing, whereas any
analysis from the expansion history would obtain a growth rate
commensurate with eq~(\ref{eq:f_w}).

\begin{figure*} 
\centering
\begin{center}
\epsfxsize=3.6in
\epsffile{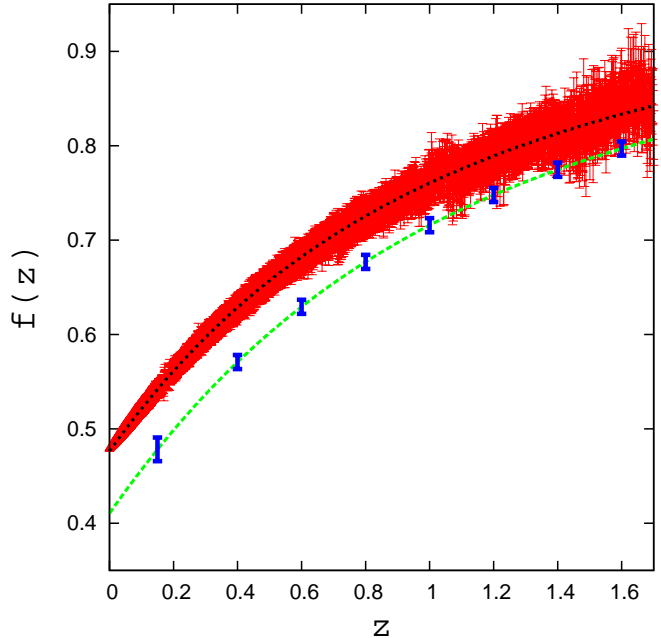} 
\end{center}
\vspace{0.0cm}
\caption{\small 
Reconstructed growth rate $f(z)$ for JDEM-like dataset using a
modified gravity model (DGP, eq~(\ref{eq:dgp})). The thick black
dotted line represents the result expected from just the expansion
history (eq~(\ref{eq:f_w}) using eq~(\ref{eq:dgp})), while the green
dashed line represents the result expected from gravitational
clustering (eq~(\ref{eq:f_dgp})). The red solid lines show the
$1\sigma$ error bars for the integral reconstruction using
eq~(\ref{eq:int}). The blue vertical lines show the expected
observational constraints from Euclid \citep{euclid}. The discrepancy
between the two would act as a signal for modified gravity.  }
\label{fig:dgp}
\end{figure*}

Therefore, if the growth rate for this model is reconstructed using on
the one hand, supernova data, and on the other, galaxy redshift
distortions, we expect the results to be different. We reconstruct the
growth rate using the JDEM-like SNe distribution for this modified
gravity model by substituting Eq~(\ref{eq:dgp}) into the integral
reconstruction method described by Eq~(\ref{eq:int}). The result is
shown in figure~\ref{fig:dgp}. For comparison, we have also plotted
the expected observational constraints from galaxy redshift
distortions for the future Euclid mission \citep{euclid}. We see that
the two results are strongly discrepant, especially at low
redshifts. If the origin of dark energy were indeed geometrical in
nature, comparisons of this sort would provide crucial evidence for
it.

Despite the popularity of the DGP model, it is currently facing
several difficulties both of an observational and theoretical nature:
Tension between this model and observational data sets has been
pointed out in \citep{dgp_obs,brane_obs,brane_obs1,brane_obs2} and the
presence of a ghost in DGP gravity \citep{ghost,ghost1,ghost2,ghost3}
may be even more problematic.  Consequently our purpose in the present
section has been to treat DGP cosmology as a toy model, used to
demonstrate the utility of the reconstruction approach developed in
this paper. (Note however the existence of other braneworld models
which are ghost free \cite{brane,brane_z2} and agree well with
observations \cite{brane_obs}.)

\subsection{Current SNe Data}

\begin{figure*} 
\centering
\begin{center}
\epsfxsize=3.6in
\epsffile{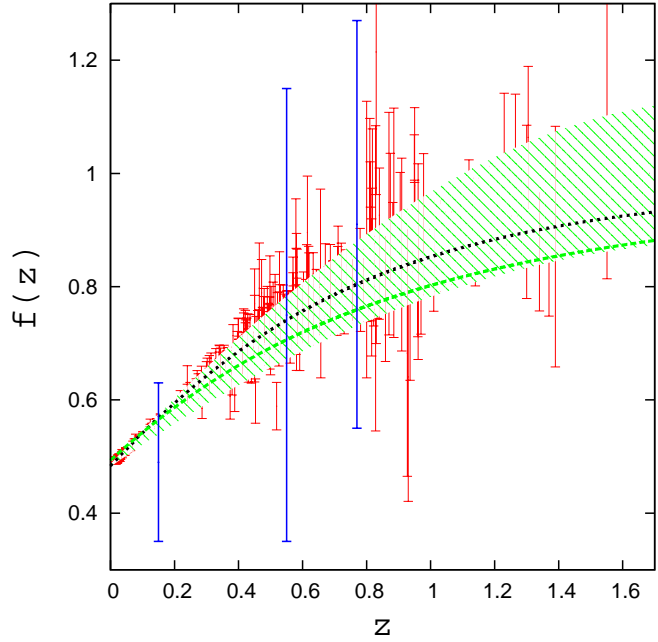} \\
\end{center}
\vspace{0.0cm}
\caption{\small 
Reconstructed growth rate $f(z)$ for current Union set of SNe data,
using $\om = 0.26 \pm 0.03$. The red solid lines show the $1\sigma$
limits for reconstructed growth parameter using the integral
reconstruction method, eq~(\ref{eq:int}), while the green dashed
shaded area shows the $1\sigma$ limits for the parameter using
smoothing scheme, eq~(\ref{eq:smooth}), for the integral
reconstruction methods. The black dotted line shows $f(z)$ for LCDM,
the green dashed line shows $f(z)$ for Model 2 (variable $w$,
eq~(\ref{eq:model2})). The three vertical blue lines show the current
measurements of $f(z)$ from 2dFGRS \citep{sdss2}, 2SLAQ \citep{slaq} and
VVDS \citep{guzo}.  }
\label{fig:real}
\end{figure*}

In figure~\ref{fig:real} we show the reconstructed growth parameters
for the currently available supernova data-- the Union dataset
\citep{union}. The results are marginalized over $\om = 0.26 \pm
0.03$, the currently accepted value of $\om$ \citep{wmap5}. The
nuisance parameter $\cal M$ which contains information on $H_0$ is
also marginalized over. For the non-smoothed method, since errors are
quite large, it is difficult to put any constraints on the growth
parameters. If the smoothing scheme is used, $f(z)$ may be constrained
to $\sim 6 \%$ at $z=0.3$. At this redshift, the growth factor $g(z)$
would be constrained to $\sim 5\%$. The reconstructed $f(z)$ is
commensurate with the cosmological constant model as well as Model 2
(variable $w$, eq~(\ref{eq:model2})) used in this paper. We also show
the three current observations of $f(z)$ from galaxy redshift-space
distortions \citep{guzo,sdss2,slaq}. The error bars on these
observations are at present quite large, but it is expected that
future data in this field would be comparable with our results from
supernovae, thus we would be able to discern physical and geometrical
DE using these different techniques (as shown in
section~\ref{dgp}). Table~\ref{tab:real} shows the $1\sigma$ limits on
the growth parameters for the reconstruction.

\subsection{Data expected from future BAO experiments}

We now check the method with BAO data. The SDSS baryon acoustic
oscillation survey of BOSS is expected to measure the baryon acoustic
oscillation power spectrum very accurately. The expected accuracy on
the angular diameter distance $d_A$ is of the order of $1.0\%$ at $z =
0.35$, $1.1\%$ at $z = 0.6$, and $1.5\%$ at $z = 2.5$, with errors on
H(z) of $1.8\%$, $1.7\%$ and $1.5\%$ at the same redshifts
\citep{sdss}. We populate a redshift range of $z=0.2-2.5$ with $20$
datapoints with errors based on these numbers and use this dataset to
reconstruct the growth parameters. Since there are only $20$ points in
the dataset, and not many at very low redshifts, the integration is
not very accurate even though the errors on $d_A$ and $H$ are
small. We find that for this dataset $g(z)$ and $f(z)$ are both
constrained to $\sim 9 \%$ at $z=2.5$ (see Tables~\ref{tab:lcdm} and
~\ref{tab:varw}, bottom row). Although these errors appear to be large
compared to those from the SNe data, for a high redshift of $z=2.5$,
these errors are actually commensurate to the errors from SNe. The
advantage of using the BAO is that we obtain the growth parameters at
a higher redshift, which is complementary to the SNe results.  In the
future, if systematics are controlled, and probes like JDEM are able
to measure both SNe and BAO data, we should be able to obtain
independent estimates of the growth parameters at both very low and
very high redshifts from this method.

\section{Conclusions}\label{concl}

In this paper we have proposed a method for extracting growth
parameters for dark energy models (within the spatially flat FRW
universe) from observations that map the background universe, such as
measures of the luminosity distance or the angular diameter
distance. The method is model independent and unbiased. For current
data, the growth factor $g(z)$ may be constrained to $\sim 5 \%$ at $z
= 0.3$, while the growth rate $f(z)$ is constrained to $\sim 6 \%$.
For future JDEM SNe data, we will be able to put constraints of the
order of a few percent on the growth parameters, \eg $2\%$ on the
growth factor and $3\%$ on the growth rate at a redshift of $0.3$ ,
and $4\%$ on the growth factor and $8\%$ on the growth rate at a
redshift of unity. In conjunction with the likelihood parameter
estimation method, this method acts as an important consistency check
on the accuracy of the priors on $\om$ for SNe. With future probes
like JDEM and BOSS taken in conjunction, it will lead to an unbiased
estimation of the growth parameters upto a redshift of $z=2.5$.

It is well known that, in GR and for most DE models, the expansion
history completely determines the linearized growth rate of density
perturbations \citep{star98,ss06} (the exact conditions for this are
formulated at the beginning of Sec. II). Consequently, a comparison of
the density contrast reconstructed from the expansion history would
provide one more important consistency check for a large variety of DE
models including the cosmological constant and quintessence. On the
other hand, as explained in more detail at the beginning of Sec. II,
any departure of the observed density contrast from that reconstructed
using standard candles and rulers would almost certainly indicate that
either there is an exchange of energy between DE and DM (so that the
effective energy-momentum tensor of DE is not on its own covariantly
conserved), or that cosmic acceleration is a consequence of modified,
non-Einsteinian gravity. In modified gravity theories, such as
Braneworld models, scalar-tensor and $f(R)$ gravity, etc., the
linearized perturbation equation for $\del$ does not follow the
Newtonian form, Eq~(\ref{eq:diff})
\citep{f_dgp,brane_pert_DGP,brane_pert,fR,beps00,jain,koy,zhao,song1}.
Hence the density contrast reconstructed using observations of
standard candles/rulers via Eq~(\ref{eq:int}) and the density contrast
determined directly from observations of large scale structure, say,
by weak lensing, galaxy redshift distortions or cluster abundances at
different $z$ \citep{guzo,percival,vikh,euclid}, are likely to differ.

In Sec.~\ref{dgp} we show that one can obtain a strong signature of
modified gravity by comparing results of this reconstruction method
with future observations of galaxy redshift distortions using the DGP
model as a toy example of modified gravity, where the growth factor
$g(z)$ is scale-independent on small scales. However, as discussed in
Sec. II, $g(z)$ often becomes scale-dependent both in modified gravity
and in the case of direct DE--DM interaction (or their
unification). Therefore, for further discrimination of DE models
alternative to quintessence and the cosmological constant, measurement
of $\del$ at different comoving scales is required to determine if
$g(z)$ is scale-dependent or not.

Future surveys such as JDEM are expected to deliver high quality data
for both supernovae and weak lensing. Using such surveys it would then
be possible to compare the reconstructed density contrast from
standard candles (SNe) with the density contrast observed from
gravitational clustering (lensing). Therefore, we hope that the
techniques developed in this paper, combined with future observations,
will help unravel the nature of that most enigmatic quantity -- dark
energy.

\section{Acknowledgements}

AAS acknowledges RESCEU hospitality as a visiting professor. He was
also partially supported by the grant RFBR 08-02-00923 and by the
Scientific Programme ``Astronomy'' of the Russian Academy of Sciences.
UA acknowledges support from the LDRD program at Los Alamos National
Laboratory and useful discussions with S. Habib, D. Holz and Z. Lukic.

\newpage

\begin{table*}
\begin{center}
{\footnotesize
\caption{\scriptsize Reconstructed linear growth factor $g$ and growth rate $f$ using different datasets for Model 1}
\begin{tabular}{c|c|ccc|ccc}
\hline
Datasets&$~~~~z~~~~$&$~~~~g(z)~~~~~$&$~~~~~g_{\rm smooth}(z)~~~~$&$~g_{\rm exact}(z)~$&$~~~~f(z)~~~~~$&$~~~~~f_{\rm smooth}(z)~~~~$&$~f_{\rm exact}(z)~$\\
\hline
&$0.3$&$1.11 \pm 0.04$&$1.12 \pm 0.02$&$1.12$&$0.65 \pm 0.04$&$0.63 \pm 0.03$&$0.64$\\
&$0.6$&$1.18 \pm 0.05$&$1.18 \pm 0.03$&$1.19$&$0.77 \pm 0.06$&$0.75 \pm 0.04$&$0.76$\\
A (Union SNe)&$1.0$&$1.27 \pm 0.06$&$1.26 \pm 0.05$&$1.25$&$0.82 \pm 0.09$&$0.83 \pm 0.06$&$0.85$\\
&$1.5$&$1.24 \pm 0.15$&$1.26 \pm 0.09$&$1.28$&$0.97 \pm 0.21$&$0.94 \pm 0.10$&$0.92$\\
\hline
&$0.3$&$1.13 \pm 0.02$&$1.12 \pm 0.01$&$1.12$&$0.64 \pm 0.02$&$0.63 \pm 0.01$&$0.64$\\
&$0.6$&$1.21 \pm 0.03$&$1.20 \pm 0.02$&$1.19$&$0.75 \pm 0.04$&$0.76 \pm 0.02$&$0.76$\\
B (JDEM SNe)&$1.0$&$1.24 \pm 0.05$&$1.23 \pm 0.03$&$1.25$&$0.86 \pm 0.07$&$0.84 \pm 0.03$&$0.85$\\
&$1.5$&$1.25 \pm 0.08$&$1.26 \pm 0.09$&$1.28$&$0.93 \pm 0.11$&$0.92 \pm 0.06$&$0.92$\\
\hline
C (BOSS BAO)&$2.5$&$1.28 \pm 0.13$&$1.29 \pm 0.11$&$1.30$&$1.01 \pm 0.09$&$1.00 \pm 0.07$&$0.97$\\
\hline
\end{tabular}\label{tab:lcdm}
}
\end{center}
\end{table*}

\begin{table*}
\begin{center}
{\footnotesize
\caption{\scriptsize Reconstructed linear growth factor $g$ and growth rate $f$ using different datasets for Model 2}
\begin{tabular}{c|c|ccc|ccc}
\hline
Datasets&$~~~~z~~~~$&$~~~~g(z)~~~~~$&$~~~~~g_{\rm smooth}(z)~~~~$&$~g_{\rm exact}(z)~$&$~~~~f(z)~~~~~$&$~~~~~f_{\rm smooth}(z)~~~~$&$~f_{\rm exact}(z)~$\\
\hline
&$0.3$&$1.12 \pm 0.03$&$1.12 \pm 0.01$&$1.13$&$0.60 \pm 0.04$&$0.61 \pm 0.03$&$0.61$\\
&$0.6$&$1.20 \pm 0.05$&$1.20 \pm 0.03$&$1.21$&$0.70 \pm 0.05$&$0.72 \pm 0.05$&$0.72$\\
A (Union SNe)&$1.0$&$1.26 \pm 0.07$&$1.27 \pm 0.05$&$1.28$&$0.82 \pm 0.09$&$0.81 \pm 0.07$&$0.80$\\
&$1.5$&$1.34 \pm 0.18$&$1.33 \pm 0.10$&$1.32$&$0.89 \pm 0.20$&$0.90 \pm 0.13$&$0.87$\\
\hline
&$0.3$&$1.13 \pm 0.02$&$1.11 \pm 0.01$&$1.13$&$0.62 \pm 0.02$&$0.61 \pm 0.01$&$0.61$\\
&$0.6$&$1.19 \pm 0.04$&$1.20 \pm 0.02$&$1.21$&$0.71 \pm 0.03$&$0.70 \pm 0.02$&$0.72$\\
B (JDEM SNe)&$1.0$&$1.27 \pm 0.05$&$1.27 \pm 0.04$&$1.28$&$0.81 \pm 0.04$&$0.80 \pm 0.03$&$0.80$\\
&$1.5$&$1.33 \pm 0.08$&$1.31 \pm 0.07$&$1.32$&$0.88 \pm 0.08$&$0.86 \pm 0.04$&$0.87$\\
\hline
C (BOSS BAO)&$2.5$&$1.37 \pm 0.13$&$1.35 \pm 0.07$&$1.34$&$0.96 \pm 0.11$&$0.94 \pm 0.06$&$0.93$\\
\hline
\end{tabular}\label{tab:varw}
}
\end{center}
\end{table*}

\begin{table*}
\begin{center}
{\footnotesize
\caption{\scriptsize Reconstructed linear growth factor $g$ and growth rate $f$ using current supernova data}
\begin{tabular}{c|ccc|ccc}
\hline
$~~~~z~~~~$&$~~~~g(z)~~~~$&$~~~~g_{\rm smooth}(z)~~~~$&$~g_{\ld {\rm CDM}}(z)~$&$~~~~f(z)~~~~$&$~~~~f_{\rm smooth}(z)~~~~$&$~f_{\ld {\rm CDM}}(z)~$\\
\hline
$0.3$&$1.13 \pm 0.05$&$1.13 \pm 0.05$&$1.12$&$0.62 \pm 0.06$&$0.61 \pm 0.04$&$0.64$\\
$0.6$&$1.21 \pm 0.07$&$1.20 \pm 0.06$&$1.19$&$0.79 \pm 0.08$&$0.80 \pm 0.05$&$0.76$\\
$1.0$&$1.29 \pm 0.10$&$1.28 \pm 0.07$&$1.25$&$0.93 \pm 0.11$&$0.88 \pm 0.08$&$0.85$\\
$1.5$&$1.37 \pm 0.16$&$1.35 \pm 0.12$&$1.28$&$1.05 \pm 0.24$&$0.98 \pm 0.11$&$0.92$\\
\hline
\end{tabular}\label{tab:real}
}
\end{center}
\end{table*}

\end{document}